\documentclass[a4paper,man,floatsintext,natbib, donotrepeattitle]{apa6}

\usepackage[english]{babel}
\usepackage[utf8x]{inputenc}
\usepackage{amsmath, amssymb, amsthm}
\usepackage{graphicx}
\usepackage[colorinlistoftodos]{todonotes}
\usepackage{hyperref}

\usepackage{mwe}
\usepackage{float}
\usepackage{subfig}
\usepackage{tikz} 
\usetikzlibrary{arrows.meta,shapes} 
\usepackage{pgf,tikz}
\usetikzlibrary{arrows,shapes.arrows,shapes.geometric,shapes.multipart,decorations.pathmorphing,positioning,shapes.swigs}
\usepackage{appendix}

\newtheorem{assumption}{Assumption}
\newtheorem{definition}{Definition}
\newtheorem{proposition}{Proposition}
\newtheorem{remark}{Remark}

\title{
{\LARGE\bfseries
Refining the Notion of No Anticipation \\
in Difference-in-Differences Studies}
}

\author{
Marco Piccininni\textsuperscript{1},
Eric J. Tchetgen Tchetgen\textsuperscript{2,3},
and Mats J. Stensrud\textsuperscript{1}
}

\affiliation{
{\footnotesize
\textsuperscript{1} Institute of Mathematics,
École Polytechnique Fédérale de Lausanne,
Lausanne, Switzerland \\[-0.3em]
\textsuperscript{2} Department of Statistics and Data Science,
The Wharton School, University of Pennsylvania, USA \\[-0.6em]
\textsuperscript{3} Department of Biostatistics,
Epidemiology and Informatics,
University of Pennsylvania, USA}}

\shorttitle{}

\fancypagestyle{titlepage}{
    \fancyhf{}
    \fancyhead[R]{\thepage}
    
}

\abstract{We clarify the interpretation of identification strategies using difference-in-differences (DiD), which are widely applied in empirical research. In particular, DiD researchers often invoke an assumption commonly referred to as the “no-anticipation assumption”. Taken literally, this assumption states that treatments in the future have no effect on outcomes in the past. Because standard causal models rely on a temporal structure in which causes precede effects, such an assumption seems to be inherently satisfied. However, most researchers attribute a different meaning to the assumption: they interpret the assumption as ensuring no effect of policy decision, e.g., immediate behavioral changes due to knowledge about the plan to implement the policy. We discuss the consequences of this mismatch, and give an alternative formulation of no anticipation that captures DiD researchers' concerns and clarifies identification arguments.}

\begin{document}
\maketitle
\section{Introduction} \label{sec:introduction}

It is often said that difference-in-differences (DiD) methods require an assumption of ``no anticipation'' \citep{roth2023trending,callaway2021did,athey2022design, renson2023transporting, zeldow2021confounding,gong2021bounds,huang2023practical,baker2025,chernozhukov2024applied, TchetgenTchetgen2023,Arkhangelsky2024, Wooldridge2023, wooldridge2025two, sun2021estimating}. This assumption is considered to be important for the validity of causal conclusions when using DiD methodologies in sociological research, see for example, \citet{Reosti2024, Schwitter2022, Elwert2023, Keskintrk2024}. 

In this work, we study the literal and practical interpretation of the no-anticipation assumption. First, in Section ``\nameref{sec:did_methodology}'' we introduce the canonical DiD setting, and show that positivity, consistency, and a parallel trends assumption are sufficient for the DiD identification argument. In Section ``\nameref{sec:no_anticipation}'', we introduce the no-anticipation assumption as it is usually formulated in the DiD literature. We show that this standard formulation simply says that the future cannot affect the past, and therefore holds trivially in canonical causal models that respect a temporal order. Most researchers, however, do not interpret the standard no-anticipation assumption literally, as is evident by many interpretations of the assumption in plain text. In Section ``\nameref{sec:better_formalization}'', we discuss these plain text interpretations, which broadly concern how knowledge of the policy future implementation can lead to behavioral changes and actions before the implementation. We then propose a different no anticipation assumption that aligns with these plain English interpretations. In Section ``\nameref{sec:why_anticipation}'' we consider an illustrative data-generating mechanism and attempt to justify researchers' concerns about anticipation. In Section ``\nameref{sec:did_multiple}'' we consider a setting with multiple time points, which is increasingly studied with DiD methods. Concluding remarks are presented in the last section. 

\section{Identification in the canonical Difference-in-Differences setting} \label{sec:did_methodology}

Consider first a canonical setting \citep{roth2023trending} with treatment and outcome observed at two time points. Suppose that we observe independent and identically distributed draws from a near-infinite superpopulation. We consider time points $k \in \{1,2\}$ and let $A_k$ indicate whether the unit implemented the treatment ($A_k = 1$) or not ($A_k = 0$) at time $k$. At time 1, no unit implements the policy. Therefore, $A_1 = 0$ for all units. Let $Y_k$ indicate the outcome at time $k$. We are interested in studying the effect of implementing the policy at time 2 on the outcomes at time 2, i.e., the effect of a policy implemented between the first and second measurements of the outcome. For this reason, the variable $Y_1$ is referred to as ``pre-treatment'' \citep{roth2023trending}. In this story there is a clear temporal order of the variables: $A_1, Y_1, A_2, Y_2$. 

Formally, we use superscripts to denote potential outcomes \citep{hernan2020causal}. Since we are only interested in the effect of intervening on $A_2$, there is no need to consider interventions on the variable $A_1$. Thus, $Y_2^{a_2 = a^*}$ is the potential outcome at time 2 had, possibly contrary to fact, $A_2$ been set to $a^*$. The conventional estimand of interest in DiD analyses is the average treatment effect among the treated \citep{roth2023trending}, which we reformulate as:

\begin{definition}
\label{def:att_a2}
$ATT_{A_2} = \mathbb{E}(Y_2^{a_2 = 1}-Y_2^{a_2 = 0} \mid A_2 = 1)$.
\end{definition}

To fix ideas, consider the effect of Medicaid expansion on insurance coverage, which was the motivating example in \citet{roth2023trending}. Let American states be the units of interest. At time 1, no unit in the population has implemented Medicaid expansion. The variable $A_2$ represents the implementation of the Medicaid expansion at time 2, while $Y_1$ and $Y_2$ represent the insurance coverage at time 1 and 2, respectively.
The $ATT_{A_2}$ represents the ``average effect of Medicaid
expansion on insurance coverage in period 2 for the states who expanded Medicaid'' \citep{roth2023trending}.

To identify the $ATT_{A_2}$, we follow the same logic of standard DiD identification arguments. We first state a simple positivity condition.

\begin{assumption}\label{ass:positivity} $
\Pr(A_2 = a^*) > 0 \quad \forall a^* \in \{0,1\}.$
\end{assumption}

In our example, Assumption \ref{ass:positivity} requires that some states implemented the policy at time 2 and others did not. We also explicitly state a causal consistency \citep{hernan2020causal} assumption.

\begin{assumption}\label{ass:consistency}
$Y_2^{a_2 = a^*} = Y_2 \quad \text{when } A_2 = a^*.$
\end{assumption}

Assumption~\ref{ass:consistency} requires that the observed insurance coverage at time 2 for a state that implemented the Medicaid expansion policy equals the insurance coverage had we intervened to implement the policy.

Finally, standard DiD involves a parallel trends condition \citep{roth2023trending}, which we write as

\begin{assumption}\label{ass:parallel}
$\mathbb{E}(Y_2^{a_2 = 0} - Y_1 \mid A_2 = 1) = \mathbb{E}(Y_2^{a_2 = 0} - Y_1 \mid A_2 = 0).$
\end{assumption}

In the words of \citet{roth2023trending}, this assumption says that ``the average change in insurance coverage for
expansion and non-expansion states would have been the same in the absence of the Medicaid expansion''.

Assumption~\ref{ass:parallel} enforces a relation between potential outcomes in those who were (factually) treated and not treated.

Under Assumptions~\ref{ass:positivity}--\ref{ass:parallel}, the $ATT_{A_2}$ is identified as
\begin{equation} \label{eq:classicdid}
ATT_{A_2} = \left\{ \mathbb{E}(Y_2 \mid A_2 = 1) - \mathbb{E}(Y_1 \mid A_2 = 1) \right\} - \left\{ \mathbb{E}(Y_2 \mid A_2 = 0) - \mathbb{E}(Y_1 \mid A_2 = 0) \right\}, 
\end{equation}
see Appendix \ref{appendix:proof_classicdid} for a proof. This identification result is essentially identical to the standard one reported in the DiD literature, see, e.g., \citet{roth2023trending}. Throughout the identification argument, we use a notation that differs slightly from that used in previous work to emphasize that it suffices to consider interventions on $A_2$ alone.

\section{No anticipation assumption and its literal interpretation} \label{sec:no_anticipation}

Assumptions~\ref{ass:positivity}--\ref{ass:parallel} are sufficient for identification of the $ATT_{A_2}$. Many authors, however, explicitly state an additional assumption: the no-anticipation, or no anticipatory effects, assumption. The no-anticipation assumption is considered important, often hidden, and suggested to be necessary for identification of the $ATT_{A_2}$ \citep{roth2023trending}. \citet{baker2025} consider the no-anticipation assumption ``crucial to the validity of the DiD estimator''. 

To be explicit, define $Y_2^{a_1 = 0, a_2 = 0}$ as the potential outcome at time 2 had we intervened to not implement the treatment at both times, and $Y_2^{a_1 = 0, a_2 = 1}$ as the potential outcome had $A_1$ and $A_2$ been fixed to 0 and 1, respectively. Some authors \citep{gong2021bounds,huang2023practical,roth2023trending, baker2025} formulate the no-anticipation assumption as: \newline

\noindent \textbf{Standard no-anticipation assumption.} $Y_1^{a_1 = 0, a_2 = 0} = Y_1^{a_1 = 0, a_2 = 1}$ when $A_2 = 1$.
\newline

Weaker versions, where the equality only holds in expectation \citep{chernozhukov2024applied} are also used. Some versions of the assumption are stated under interventions on $A_2$ only, i.e., corresponding to $Y_1^{a_2 = 0} = Y_1^{a_2 = 1}$\citep{TchetgenTchetgen2023,Wooldridge2023}. 

Taken literally, the standard no-anticipation assumption implies that we must explicitly rule out the possibility that an intervention at a later time could affect outcomes at earlier times, a reading that makes the assumption vacuous in any causal model with a temporal order. A violation of the standard no-anticipation assumption implies that intervening at time 2 affects outcomes at time 1. This would mean, for example, that implementing Medicaid expansion in year 2 can affect the insurance coverage in year 1.

Several authors state the standard no-anticipation assumption, or similar versions adapted to DiD with multiple time periods \citep{roth2023trending,callaway2021did,athey2022design,baker2025,Arkhangelsky2024, wooldridge2025two, sun2021estimating}, which we discuss in a later section. Usually the no-anticipation assumption is interpreted differently (see Section \nameref{sec:better_formalization}), but in some passages it seems to be interpreted literally. For example, the no-anticipation assumption has been described as the assumption that ``post-treatment status should have no impact on pre-treatment outcomes'' \citep{gong2021bounds}, or that ``treatments in the “future” cannot affect outcomes observed in the “past.”'' \citep{strezhnev2024group}. The standard no-anticipation assumption in expectation is sometimes called ``NEPT'' \citep{lechner2010estimation,pronti2023water}, explained in plain English as the absence of a treatment effect on pre-treatment outcomes. For example, \citet{lechner2010estimation} describes this assumption as ``the assumption that in the pre-treatment period the treatment had no effect on the pre-treatment population''. Because it prevents the future from affecting the past, the standard no-anticipation assumption has even been denoted as a condition about the ``arrow of time'' \citep{hettinger2024framework,diffindiff_resource}, referring to the notion from classical physics that, macroscopically, time is perceived to flow in a one-way direction \citep{wiki2025arrow}.

Yet, conventional causal models are usually defined with respect to a temporal (or topological) order, where events in the future cannot affect the past. In conventional causal models, like Robins' Finest Fully Randomized Causally Interpreted Structured Tree Graphs (FFRCISTG) \citep{robins1986new}, which covers Pearl's non-parametric structural equation model with independent errors \citep{pearl2009causality} as a strict submodel, future events cannot affect the past. Richardson and Robins encode in their ``Counterfactual Existence Assumption'' the principle that the past cannot be affected by the future in FFRCISTG models \citep{richardson2013swig}. The ``Counterfactual Existence Assumption'' assumption combines the assumptions known in the literature as ``consistency'' and ``recursive substitution'' \citep{richardson2013swig}. Consistency alone, without invoking a recursive causal model, is not sufficient to prevent the future from affecting the past. However, consistency and the idea that the future does not affect the past are related, in the sense that both can be derived from the definition of a set of recursive structural equations \citep{Pearl2010}. Perhaps for this reason, some introductory texts argue that the no-anticipation assumption is implied by the equality $Y_1^{a_2 = 0} = Y_1$ \citep{callaway2024multiperoid}, corresponding to a consistency assumption for the outcome at time 1. Similarly, other authors have defined no anticipation as the equality between the observed outcome at time 1 and the counterfactual outcome at time 1 under any treatment, i.e., $Y_1^{a_2 = 0} = Y_1^{a_2 = 1} = Y_1$, even if the first outcome measurement is instantiated before the treatment \citep{renson2023transporting,zeldow2021confounding}. 

From the arguments in this section, we conclude that the standard no anticipation assumption is unnecessary when one is committed to canonical causal models. Perhaps for this reason, other works on DiD do not explicitly state the no-anticipation assumption \citep{abadie2005semiparametric,angrist2009mostly, Angrist1999, Caniglia2020, Meyer1995}.

\section{A formalization of no anticipation that corresponds to researchers' concerns} \label{sec:better_formalization}

If researchers were worried about the future affecting the past, it would be unclear why they would only preclude the effect of $A_2$ on $Y_1$, and not also the effect of $A_2$ on other events in the past. The special emphasis on the first outcome measurement suggests a more specific concern. In fact, most researchers using and studying DiD do not interpret the standard no-anticipation assumption literally.

For example, \citet{roth2023trending} explain the no-anticipation assumption by stating that ``units do not act on the knowledge of their future treatment date before treatment starts'', and they introduce the standard no-anticipation assumption to ensure that ``in years prior to Medicaid expansion, insurance coverage in states that expanded Medicaid was not affected by the upcoming Medicaid expansion''. In a related application, \citet{baker2025} invoked the standard no-anticipation assumption to prevent that announcement of Medicaid expansion had an effect on the outcome before its actual expansion. The concern here is not that the future can affect the past, but, rather, that the decision of expanding Medicaid can affect the insurance coverage before the Medicaid expansion is actually implemented; it is the knowledge of the treatment, rather than receiving the treatment itself, that affects outcomes before treatment implementation. For example, \citet{Peng2017} found evidence consistent with the fact that insurers in Pennsylvania changed premiums ``in anticipation'', that is, before the federal government approved the Medicaid expansion. This could happen, according to \citet{Peng2017}, because Pennsylvania presented the waiver application before the rates were due.

Because many policy decisions are public and known in advance, units adjust their behavior based on this knowledge, even before the policy itself is implemented. This type of anticipation is a well-known problem when estimating causal effects based on temporal data \citep{Malani2015}.

To formally represent this plain language notion of no anticipation, we introduce a variable $P$, representing the decision (or plan) of implementing the policy. To simplify the presentation, let $P = 1$ if there is a plan to implement the policy at time 2, and $P = 0$ otherwise. In the Medicaid example, a researcher could conceptualize $P$ as indicating whether the state submitted a waiver application to expand Medicaid. 

In certain settings, the random variable $P$ temporally precedes the first measurement of the outcome. In the Medicaid DiD study, for example, this happens if $Y_1$ represents a measurement of insurance coverage for a period of time following the waiver application ($P$). We will focus on settings in which we can conceive the existence of a decision variable $P$, temporally ordered before $Y_1$. Specifically, we consider the temporal order: $A_1, P, Y_1, A_2, Y_2$. Yet there exist settings where $P$ is implausible or redundant. For example, when the treatment of interest is a viral outbreak \citep{park2022} or when there is no temporal lag between the treatment decision and treatment implementation, the distinction between $P$ and $A_2$ becomes unnecessarily convoluted. However, the variable $P$ is particularly relevant in settings where authors are concerned about anticipation, for example in DiD studies evaluating the effects of new policies, like Medicaid expansion, that are typically announced prior to their implementation.

Now consider an alternative definition of no anticipation that explicitly incorporates the variable $P$, representing the decision.
\begin{assumption}
\label{ass:noanticipationP} $\mathbb{E}(Y_1^{p = 1} \mid P = 1) = \mathbb{E}(Y_1^{p = 0} \mid P = 1)$.
\end{assumption}
Here, $Y_k^{p=1}$ is the potential outcome at time $k$ had we intervened to set $P=1$.
Assumption \ref{ass:noanticipationP} states that there is no average effect of the decision $P$ on the first outcome measurement $Y_1$, for units that decided to implement the policy ($P = 1$). It is also possible to formulate a stronger version of Assumption \ref{ass:noanticipationP}, requiring that the equality of potential outcomes holds for every unit, i.e., $Y_1^{p=1}=Y_1^{p=0}$.

In the example on Medicaid, Assumption \ref{ass:noanticipationP} requires that the plan to implement the Medicaid expansion ($P$) does not affect the insurance coverage in the years before the expansion is actually implemented ($Y_1$), for states that planned the expansion. Therefore, for example, if the waiver application changes the behavior on the insurance market, Assumption \ref{ass:noanticipationP} can be violated.

We believe that the standard no-anticipation equation does not encode the conditions that the researchers intend to formalize. The standard no-anticipation assumption rules out that the future treatment itself affects past outcomes. This assumption holds even when knowledge of the upcoming policy affects units' behaviors. On the other hand, the new formulation based on a treatment plan better captures researchers concerns. Assumption \ref{ass:noanticipationP} precludes that the \textit{plan} of implementing the treatment affects outcomes in the pretreatment period. In particular, Assumption \ref{ass:noanticipationP} is non-trivial in Robins's FFRCISTG model. 

Relatedly, \citet{Abbring2003} write that Robins's causal model relies on a consistency assumption ensuring that future treatments cannot affect the past. However, \citet{Abbring2003} state that this assumption, standard in biostatistics, is not always suitable for social sciences, as ``in social sciences, individuals may act on knowledge of the moment of realization of a future treatment''. They therefore propose a no-anticipation assumption \citep{Abbring2003} on the hazard scale, conceptually similar to the standard no-anticipation assumption. The concern, stated in plain English, is that units act on existing \textit{knowledge}; however, the formal assumption, stated in mathematical terms, encodes the impossibility of future treatment to affect the past.

These considerations and examples suggest that the investigators intend to encode assumptions that concern the decision variable $P$, not the treatment itself.

\section{Why be concerned about anticipation?} \label{sec:why_anticipation}

Anticipation, in the sense of $P$ affecting $Y_1$, does not necessarily violate assumptions \ref{ass:positivity}, \ref{ass:consistency}, or \ref{ass:parallel} that are sufficient to identify the $ATT_{A_2}$ by the DiD functional, see also example in Appendix \ref{appendix:cars}. Therefore, it would be helpful to clarify exactly the reasons why researchers are worried about anticipation and how anticipation violates their identification conditions. 

To illustrate possible explanations for their concerns, we consider a case where the decision to implement the policy determines inevitably that the policy will be implemented. In this case, there are at least two reasons to be concerned about anticipation. One reason is that researchers interpret the parallel trends assumption with respect to the intervention setting $P=0$, rather than $A_2=0$, and their estimand of interest is $ATT_{A_2}$. Another reason is that they interpret the parallel trends assumption with respect to $P=0$, but their estimand of interest is an effect defined by interventions on $P$, not on $A_2$.

\subsection{Setup}

We assume that $P$ deterministically causes $A_2$:
\begin{assumption}
\label{ass:determinism2times}
$A_2^{p=p^*} = p^*$ for every $p^* \in \{0,1\}$.
\end{assumption}
In words, when a decision about the policy is made, the decision is always executed. In the Medicaid example, if we interpret $P$ as the waiver application, Assumption \ref{ass:determinism2times} states that Medicaid expansion occurs if and only if the state submits a waiver application: if the application is submitted, the expansion is always approved; if it is not submitted, the expansion does not occur. Assumption \ref{ass:determinism2times} is reasonable when $P$ represents the decision of a parliament or legislative body in the region, and the decision is to implement a policy in the next years or starting at a specific date in the future. This determinism does not always hold; it can happen, for example, that a policy is announced but not implemented. However, this determinism is not needed for the identification result presented in Section ``\nameref{sec:did_methodology}''. Conceptually, the parallel trends assumption makes it possible to identify treatment effect even in the presence of unmeasured confounding. However, $P$ can be regarded as one such unmeasured confounder. Thus, if the parallel trends assumption \ref{ass:parallel} holds, even when the plan $P$ or other variables, such as individuals’
beliefs, affect future treatment implementation and outcomes, the identification results will be valid. Here, we give an attempt to justify why researchers are concerned about anticipatory behaviors when using classic DiD functionals.

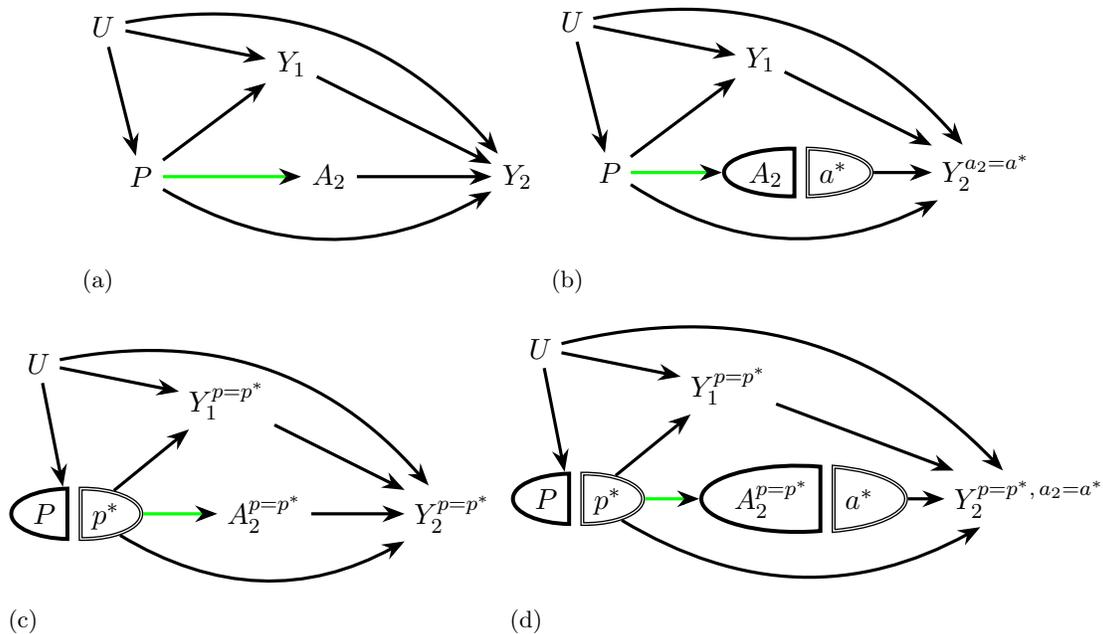
\begin{figure}[H]
\begin{center}
\subfloat[]
        \centering
        \begin{tikzpicture}
            \begin{scope}[every node/.style={thick,draw=none}]
            \node[name=P] at (-3,0){$P$};
            \node[name=A2] at (-0.5,0){$A_2$};
            \node[name=Y1] at (-1,1.5){$Y_1$};
            \node[name=Y2] at (2,0){$Y_2$};
            \node[name=U] at (-3.5,2){$U$};
        \end{scope}
        
        \begin{scope}[>={Stealth[black]},
                      every node/.style={fill=white,circle},
                      every edge/.style={draw=black,very thick}]
            \path[->] (U) edge (P);
            \path[->] (U) edge[bend left] (Y2);      
            \path[->] (U) edge (Y1);
            \path[->] (P) edge (Y1);
            \path[->] (P) edge[green] (A2);
            \path[->] (P) edge[bend right] (Y2);
            \path[->] (Y1) edge (Y2);
            \path[->] (A2) edge (Y2);
        \end{scope}
        \end{tikzpicture}
\subfloat[]
        \centering
        \begin{tikzpicture}
        \tikzset{line width=1.5pt, outer sep=0pt,
        ell/.style={draw,fill=white, inner sep=2pt,
        line width=1.5pt},
        swig vsplit={gap=5pt,
        inner line width right=0.5pt},
        swig hsplit={gap=5pt}
        };
            \node[name=A2,shape=swig vsplit] at (-0.5,0) {
                                              \nodepart{left}{$A_2$}
                                              \nodepart{right}{$a^*$} };
                        \node[name=P] at (-3,0){$P$};
             \node[name=Y1] at (-1,1.5){$Y_1$};
            \node[name=Y2] at (2,0){$Y_2^{a_2=a^*}$};
            \node[name=U] at (-3.5,2){$U$};
        \begin{scope}[>={Stealth[black]},
                      every edge/.style={draw=black,very thick}]
            \path[->] (U) edge (P);
            \path[->] (U) edge[bend left] (Y2);          
            \path[->] (U) edge (Y1);
            \path[->] (P) edge (Y1);
            \path[->] (P) edge[green] (A2);
            \path[->] (P) edge[bend right] (Y2);
            \path[->] (Y1) edge (Y2);
            \path[->] (A2) edge (Y2);
        \end{scope}
        \end{tikzpicture}
\subfloat[]
        \centering
        \begin{tikzpicture}
        \tikzset{line width=1.5pt, outer sep=0pt,
        ell/.style={draw,fill=white, inner sep=2pt,
        line width=1.5pt},
        swig vsplit={gap=5pt,
        inner line width right=0.5pt},
        swig hsplit={gap=5pt}
        };
            \node[name=A2] at (-0.5,0){$A_2^{p=p^*}$};
            \node[name=P,shape=swig vsplit] at (-3,0) {
                                              \nodepart{left}{$P$}
                                              \nodepart{right}{$p^*$} };
             \node[name=Y1] at (-1,1.5){$Y_1^{p=p^*}$};
            \node[name=Y2] at (2,0){$Y_2^{p=p^*}$};
            \node[name=U] at (-3.5,2){$U$};
        \begin{scope}[>={Stealth[black]},
                      every edge/.style={draw=black,very thick}]
            \path[->] (U) edge (P);
            \path[->] (U) edge[bend left] (Y2);        
            \path[->] (U) edge (Y1);
            \path[->] (P) edge (Y1);
            \path[->] (P) edge[green] (A2);
            \path[->] (P) edge[bend right] (Y2);
            \path[->] (Y1) edge (Y2);
            \path[->] (A2) edge (Y2);
        \end{scope}
        \end{tikzpicture}
\subfloat[]
        \centering
        \begin{tikzpicture}
        \tikzset{line width=1.5pt, outer sep=0pt,
        ell/.style={draw,fill=white, inner sep=2pt,
        line width=1.5pt},
        swig vsplit={gap=5pt,
        inner line width right=0.5pt},
        swig hsplit={gap=5pt}
        };
            \node[name=A2,shape=swig vsplit] at (0,0) {
                                              \nodepart{left}{$A_2^{p=p^*}$}
                                              \nodepart{right}{$a^*$} };
            \node[name=P,shape=swig vsplit] at (-3,0) {
                                              \nodepart{left}{$P$}
                                              \nodepart{right}{$p^*$} };
             \node[name=Y1] at (-1,1.5){$Y_1^{p=p^*}$};
            \node[name=Y2] at (3,0){$Y_2^{p=p^{*},\, a_2=a^*}$};
            \node[name=U] at (-3.5,2){$U$};
        \begin{scope}[>={Stealth[black]},
                      every edge/.style={draw=black,very thick}]
            \path[->] (U) edge (P);
            \path[->] (U) edge[bend left] (Y2);        
            \path[->] (U) edge (Y1);
            \path[->] (P) edge (Y1);
            \path[->] (P) edge[green] (A2);
            \path[->] (P) edge[bend right] (Y2);
            \path[->] (Y1) edge (Y2);
            \path[->] (A2) edge (Y2);
        \end{scope}
        \end{tikzpicture}
\end{center}
\caption{Panel (a) is the directed acyclic graph representing the assumed causal relationships between the decision to implement the policy ($P$), the policy implementation at time 2 ($A_2$), the outcome measurements at time 1 and 2 ($Y_1$ and $Y_2$), and possible unmeasured common causes ($U$). Panel (b) represents the single world intervention graph corresponding to the directed acyclic graph in Panel (a) when intervening on the policy implementation at time 2, setting $A_2 = a^*$. Panel (c) represents the single world intervention graph when intervening on the decision to implement the policy, setting $P = p^*$. Panel (d) represents the single world intervention graph when intervening on both $P$, setting $P=p^*$, and on $A_2^{p=p^*}$. A deterministic relationship (in green) is assumed between $P$ and $A_2$.}
 \label{fig:causal_graphs}
\end{figure}

Let $U$ represent possibly unmeasured common causes of $P,\, Y_1$, and $Y_2$. We can encode the causal structure, including $P$, in a causal directed acyclic graph representing a FFRCISTG model \citep{robins1986new,richardson2013swig} (Figure~\ref{fig:causal_graphs}a). 
This causal model makes explicit that we can conceive of two possible distinct interventions: interventions on $P$, the decision to implement the policy; or on $A_2$, the policy implementation itself. We represented the intervention on the variable $A_2$ in Figure \ref{fig:causal_graphs}b, on the variable $P$ in Figure \ref{fig:causal_graphs}c, and on both $A_2$ and $P$ in Figure \ref{fig:causal_graphs}d, using single world intervention graphs \citep{richardson2013swig}.

\begin{remark}[Relation to Separable Effects]
    The interventions on $P$ and $A_2$ are conceptually similar to interventions encountered in separable effect settings \citep{stensrud2022conditional, stensrud2021generalized, martinussen2023estimation, richardson2010alternative, richardson2020interventionist, wen2024causal,park2024proximal}. That is, in the observed data, $A_2$ and $P$ are deterministically related (Assumption \ref{ass:determinism2times}), but we consider hypothetical interventions on them, so that they might take different values (Figure \ref{fig:causal_graphs}d). These estimands are single-world quantities \citep{richardson2013swig}, that could, in principle, be realized in a future experiment. That is, the results we present are, in principle, falsifiable. For example, we could imagine an experiment where the Medicaid expansion occurs independently of the state’s decision to apply for a waiver, or in which the expansion is blocked despite the state having applied for it.
\end{remark}

\subsection{Belief in a different parallel trends assumption} \label{sec:different_parallel}

Assumption \ref{ass:parallel} reflects the classical parallel trends assumption \citep{roth2023trending}. Consider now a parallel trends assumption with respect to $P$, rather than $A_2$:
\begin{assumption}
\label{ass:paralleltrendP}
$\mathbb{E}(Y_2^{p = 0} - Y_1^{p = 0} \mid P = 1) = \mathbb{E}(Y_2^{p = 0} - Y_1^{p = 0} \mid P = 0)$.
\end{assumption}
Assumption \ref{ass:paralleltrendP} states that the trend would have been parallel in units with observed $P = 1$ and $P = 0$, had we forced a decision of not implementing the policy. Under the determinism in Assumption \ref{ass:determinism2times}, Assumption \ref{ass:paralleltrendP} means that the trends would have been parallel if the treatment group never \textit{decided} to implement the policy. This interpretation differs from the parallel trends in Assumption \ref{ass:parallel}. For example, according to Assumption \ref{ass:paralleltrendP} states that did not apply for the Medicaid waiver and those that did should have followed the same trend in insurance coverage, had the latter not applied. 

Consider also the following consistency assumption, concerning interventions on $P$.
\begin{assumption}
\label{ass:consistencyP}
$Y_1^{p = p^*} = Y_1,\, Y_2^{p = p^*} = Y_2,$ and $A_2^{p = p^*} = A_2$ when $P = p^*$.
\end{assumption}

Next, suppose that the decision $P$ only affects $Y_2$ through the mediator $A_2$. 
To formalize this exclusion restriction, we consider a joint intervention on the policy decision and the policy implementation, and use $Y^{p=p^{*},\, a_2=a^*}_2$ to indicate the outcome at time 2 when we set $P=p^*$ and $A^{p=p^*}_2=a^*$. This joint intervention is represented in the single world intervention graph in Figure \ref{fig:causal_graphs}d and the exclusion restriction is formalized as follows:
\begin{assumption} \label{ass:no_directeffectP}
    $\mathbb{E}(Y_2^{p = p^*,a_2=a^*} \mid P=p')=\mathbb{E}(Y_2^{a_2=a^*} \mid P=p')$ for every $p^*,a^*,p' \in \{0,1\}$.
\end{assumption}
Assumption \ref{ass:no_directeffectP} states that the decision to implement a Medicaid expansion does not affect insurance coverage at time 2, other than through the Medicaid expansion itself. Consider a case where Assumption \ref{ass:no_directeffectP} fails. Suppose that we are interested in the effect of a policy on GDP. When the decision to implement the policy is made public, it sparks a wave of major strikes and protests. In this case, the effect of the decision is not fully mediated by the policy implementation, as the strikes have long-term consequences and affect economic activity and the GDP in the future.

By explicitly introducing $P$, representing the policy decision, we clarify that it is possible to justify the concern about anticipatory behavior when researchers believe in a parallel trends assumption with respect to $P=0$:

\begin{proposition} \label{prop:same_parallel_trend}
Suppose that Assumptions \ref{ass:positivity}, \ref{ass:determinism2times}, \ref{ass:paralleltrendP}, \ref{ass:consistencyP}, and \ref{ass:no_directeffectP} hold. Then,
\[\left\{ \mathbb{E}(Y_2 \mid A_2=1) - \mathbb{E}(Y_1 \mid A_2=1) \right\} - \left\{ \mathbb{E}(Y_2 \mid A_2=0) - \mathbb{E}(Y_1 \mid A_2=0) \right\} = ATT_{A_2} - \psi    
\]
where
\[\psi= \mathbb{E}(Y^{p=1}_1 \mid P=1) - \mathbb{E}(Y_1^{p = 0} \mid P = 1).
\]
\end{proposition}
See Appendix \ref{appendix:different_parallel_trend} for a proof. Conceptually, suppose that the effect of $P$ on $Y_2$ is entirely mediated by $A_2$, and $P$ and $A_2$ are deterministically related. Then, under the parallel trends with respect to $P$, the classic DiD functional in Equation \ref{eq:classicdid} is equal to the $ATT_{A_2}$ minus a bias term $\psi$. This bias term $\psi$ is exactly the anticipation effect, that is, the effect of $P$ on $Y_1$ among units that planned the treatment. Therefore, when there is no anticipation (Assumption \ref{ass:noanticipationP} holds), $\psi=0$ and the classic DiD functional identifies the $ATT_{A_2}$.

The concern that units adjust their behavior in response to the policy plan is reasonable if researchers believe in Assumption~\ref{ass:paralleltrendP}. Proposition \ref{prop:same_parallel_trend} says that, under the stated assumptions, the classic DiD functional captures two different effects: the effect of interest ($ATT_{A_2}$), and the anticipation effect ($\psi$). This result aligns with informal statements from the literature, for example, \citet{roth2023trending} state that if anticipation is present ``changes in the outcome for the treated group between period 1 and 2 could reflect not just the
causal effect in period t = 2 but also the anticipatory effect in period t = 1'' . This quote is compatible with the idea that researchers, perhaps implicitly, interpret the parallel trends assumption under intervention $P = 0$ (Assumption~\ref{ass:paralleltrendP}), while instead writing the assumption under $A_2 = 0$ (Assumption~\ref{ass:parallel}).

In the next section, we discuss another possible reason for being concerned about anticipation.

\subsection{Interest in a different effect} \label{sec:different_intervention}

The $ATT_{A_2}$ represents the effect of intervening on $A_2$, among units who naturally would be treated (observed to have $A_2 = 1$). This effect can be different from an effect of the policy decision, i.e., intervening to change the value of $P$. The single world intervention graphs in Figure~\ref{fig:causal_graphs}b and \ref{fig:causal_graphs}c represent the two hypothetical interventions.

To formalize the intervention on the decision, consider the following estimand:

\begin{definition}\label{def:att_p}
$ATT_P = \mathbb{E}(Y_2^{p = 1} - Y_2^{p = 0} \mid P = 1).$
\end{definition}
The $ATT_P$ represents the average treatment effect among the treated, where the treatment is now the variable $P$ and not, as previously considered, $A_2$. In the Medicaid example, the $ATT_P$ would correspond to the effect of deciding to expand Medicaid among states that in fact decided to expand. To further highlight the conceptual difference between $ATT_P$ and $ATT_{A_2}$, we present a didactic example in Appendix \ref{appendix:cars}.

Some researchers seem to be interested in the $ATT_P$, rather than the $ATT_{A_2}$, when they believe anticipation is present. For example, \citet{baker2025} noted that the announcement of Medicaid expansion could affect their outcome before the policy was actually rolled out. They claimed that in this case the ``treatment'' starts at the time of the announcement, not at the time of implementation \citep{baker2025}. Thus, here the treatment seems to be the decision of expanding Medicaid, not the expansion itself, which occurred later. Similarly, others have proposed to redefine treatment as the announcement of a decision, in order to ``immunize'' the DiD analyses against anticipation effects \citep{deChaisemartin2022}. Defining treatment at the time a policy is announced is a common approach to circumvent issues related to anticipation \citep{Cerulli2022}.

In the setting with two time points, when the no-anticipation Assumption \ref{ass:noanticipationP} holds and $P$ is the treatment of interest, $ATT_P$ is identified by the classical DiD functional. We clarify this in the following proposition:
\begin{proposition}
\label{prop:didwithp_2times}
Under Assumptions~\ref{ass:positivity}, \ref{ass:noanticipationP}, \ref{ass:determinism2times}, \ref{ass:paralleltrendP}, and \ref{ass:consistencyP} 
\[
ATT_P = \left\{ \mathbb{E}(Y_2 \mid A_2 = 1) - \mathbb{E}(Y_1 \mid A_2 = 1) \right\} - \left\{ \mathbb{E}(Y_2 \mid A_2 = 0) - \mathbb{E}(Y_1 \mid A_2 = 0) \right\}.
\]
\end{proposition}
See Appendix \ref{appendix:didwithp_2times} for details about the proof. 

Informally, classical DiD methods target the $ATT_P$ if there is no anticipation, $A_2$ is deterministically equal to $P$, and parallel trends with respect to an intervention fixing $P=0$ holds. Under the assumptions invoked in Proposition \ref{prop:didwithp_2times}, the $ATT_{A_2}$ and the $ATT_{P}$ can still differ. However, when the determinism in Assumption \ref{ass:determinism2times} and the full-mediation condition in Assumption \ref{ass:no_directeffectP} hold, the two estimands coincide. 

As discussed in previous sections, the interventions we consider on $P$ and $A_2$ are related to the interventions defining separable effects \citep{stensrud2022conditional, stensrud2021generalized, martinussen2023estimation, richardson2010alternative, richardson2020interventionist, wen2024causal,park2024proximal}. However, our identification arguments differ from those previously invoked in separable effect settings. This is because there is unmeasured confounding between $P$ and $(Y_1, Y_2)$, violating the classical separable effects conditions \citep{richardson2020interventionist,stensrud2021generalized}. Also unlike the settings in \citet{wen2024causal} and \citet{park2024proximal}, we do not have a perfect frontdoor variable or proxy variables. Instead, we rely on the parallel trends assumptions for identification.

\subsection{Extension to the setting with multiple time points} 
\label{sec:did_multiple}

A popular extension of the DiD methodology allows for staggered policy adoption and multiple outcome measurements \citep{roth2023trending,callaway2021did}. In this setting, a ``limited treatment anticipation'' assumption \citep{callaway2021did} or a no-anticipation assumption \citep{roth2023trending} are often invoked.

We will specifically consider the DiD setting from \citet{callaway2021did}, with measurements of treatments and outcomes over time. To make the exposition as simple as possible, we discuss identification arguments without conditioning on baseline covariates. However, the results can be readily adapted to the settings where DiD assumptions are conditional on baseline covariates.

As in the canonical setting, we observe units that are independent and identically distributed draws from a near-infinite superpopulation. We consider time points $k \in \{1, \ldots, \tau\}$ and indicate with $A_k$ whether the unit implemented the treatment ($A_k = 1$) or not ($A_k = 0$) at time $k$. Further, we indicate with $\bar{A}_k$ the vector of all treatment measurements from time 1 to time $k$. The outcome at time $k$ is denoted by $Y_k$. We also define the potential outcome $Y_k^{\bar{a}_\tau}$, corresponding to the value of the outcome at time $k$ if the treatment was fixed to be $\bar{A}_\tau = \bar{a}_\tau$. We use $\bar{x}_d$ with $x \in \{0,1\}$ to indicate a vector of length $d$ with elements all equal to $x$. For example, we use $Y_k^{\bar{a}_\tau = \bar{0}_\tau}$ to indicate the outcome at time $k$ had the policy never been implemented until the end of the study.

A common estimand of interest in DiD analyses with multiple time points is the ``group-time average treatment effect'' \citep{callaway2021did}:
\begin{definition} \label{def:att_Amulttime}
\[
ATT_{\bar{A}}(g, k) = \mathbb{E}\left( Y_k^{\bar{a}_\tau = \{ \bar{0}_{g-1}, \bar{1}_{\tau - g + 1} \}} - Y_k^{\bar{a}_\tau = \bar{0}_\tau} \mid \bar{A} = \{ \bar{0}_{g-1}, \bar{1}_{\tau - g + 1} \} \right)
\]
where $g \leq \tau$, $k \leq \tau$.
\end{definition}
This effect contrasts the expected outcomes at time $k$ among units that start treatment at time $g$ between two strategies: treat from time $g$ onwards, and never treat.

To identify this estimand, \citet{callaway2021did} invoked a ``limited treatment anticipation assumption'', imposing that future interventions do not affect measurements before a certain point in time in the past:
\newline

\noindent \textbf{Standard limited treatment anticipation assumption.} There is a known $\delta \geq 0$, such that
\[
\mathbb{E}(Y_k^{\bar{a}_\tau = \{ \bar{0}_{g-1}, \bar{1}_{\tau - g + 1} \}} \mid \bar{A}_\tau = \{ \bar{0}_{g-1}, \bar{1}_{\tau - g + 1} \}) = \mathbb{E}(Y_k^{\bar{a}_\tau = \bar{0}_\tau} \mid \bar{A}_\tau = \{ \bar{0}_{g-1}, \bar{1}_{\tau - g + 1} \})
\]
for all $g, k$ such that $k < g - \delta$.\newline

For example, if $\delta = 1$ this assumption allows that an intervention at time $g$ affects the outcome at time $g-1$. In other words, events in the future can affect events in the (recent) past.

While we ground our presentation on the important work of \citet{callaway2021did}, other definitions of no anticipation also exist for the multiple time periods case:\newline

\noindent \textbf{Standard staggered no-anticipation assumption.} $Y_k^{\bar{a}_\tau = \{ \bar{0}_{g-1}, \bar{1}_{\tau - g + 1} \}} = Y_k^{\bar{a}_\tau = \bar{0}_\tau}$ when $k < g$.\newline

The Standard staggered no-anticipation assumption is often used in the literature \citep{roth2023trending,athey2022design,baker2025}. The assumption states that, before time $g$, a unit that is forced to start treatment at time $g$ has the same outcome that it would have had if it was forced to never start treatment; stating that interventions in the future cannot modify outcomes in the past. Sometimes the equality is only assumed in expectation in certain subpopulations \citep{wooldridge2025two, sun2021estimating}. A stronger version equating potential outcomes at time $k$ for all possible interventions that do not differ until time $k$ also exists \citep{Arkhangelsky2024}.

Like the standard no-anticipation assumption for two time periods, the standard limited treatment anticipation and staggered no-anticipation assumptions are statements preventing the future from affecting the far past. For the same reasons discussed in Section ``\nameref{sec:better_formalization}'', these statements do not match the substantial interpretation attributed to them.

In Appendix \ref{appendix:didwithp_multtime}, we extend the results of Proposition \ref{prop:didwithp_2times} to multiple time points. Specifically, we introduce a time-varying variable representing the decision to implement the policy, enforce a determinism between the decision and the lagged treatment implementation, and show that, under no-anticipation and parallel trends assumptions conceptually similar to the ones in Assumption \ref{ass:noanticipationP} and \ref{ass:paralleltrendP}, the group-time average treatment effect for the decision is identified by a functional identical to the one in \citet{callaway2021did,callaway2024multiperoid} evaluated at a different intervention time.

\section{Conclusions}

We have refined the definition of no anticipation. Specifically, we have highlighted a mismatch between the standard mathematical formulation of the no-anticipation assumption and the plain language description of its meaning. The mismatch can be avoided by clarifying whether the interventions being considered concern the implementation of the policy or the earlier decision to implement it. Thus, we suggest explicitly including the decision variable in the causal model. This variable can then be used to define the estimand of interest and the assumptions needed for identification. Clarifying these points is important for practitioners who assess the validity of a causal analysis, which requires a critical assessment of the identification assumptions. The new formalizations could also motivate future work on identification in the presence of anticipatory effects.

Finally, while we have focused on difference-in-differences methodology, the no-anticipation assumption is not unique to this design. For example, the standard staggered no-anticipation assumption has been invoked in the context of synthetic controls \citep{BenMichael2021}. No anticipation is considered a crucial assumption in methodologies leveraging time variation \citep{Abadie2021}.

\section*{Acknowledgments}
Marco Piccininni and Mats J. Stensrud were supported by the Swiss National Science Foundation (project funding, grant number: 207436). During the preparation of this manuscript the authors used chatGPT in order to assist with formatting, minor language editing, and error checking. The authors reviewed and edited the content as needed and take full responsibility for the content of the publication.

\bibliography{references}

\appendix

\renewcommand\thesection{Appendix \Alph{section}}

\section{}
\label{appendix:proof_classicdid}

\begin{align*}
ATT_{A_2} 
&\overset{\text{(D\ref{def:att_a2})}}{=} \mathbb{E}(Y_2^{a_2=1} - Y_2^{a_2=0} \mid A_2=1) \\
&\overset{\text{(A\ref{ass:parallel})}}{=} \mathbb{E}(Y_2^{a_2=1} - Y_2^{a_2=0} \mid A_2=1) + \mathbb{E}(Y_2^{a_2=0} - Y_1 \mid A_2=1) - \mathbb{E}(Y_2^{a_2=0} - Y_1 \mid A_2=0) \\
&= \mathbb{E}(Y_2^{a_2=1} \mid A_2=1) - \mathbb{E}(Y_2^{a_2=0} \mid A_2=1) + \mathbb{E}(Y_2^{a_2=0} \mid A_2=1) - \mathbb{E}(Y_1 \mid A_2=1) \\
&\quad - \mathbb{E}(Y_2^{a_2=0} \mid A_2=0) + \mathbb{E}(Y_1 \mid A_2=0) \\
&= \mathbb{E}(Y_2^{a_2=1} \mid A_2=1) - \mathbb{E}(Y_1 \mid A_2=1) - \mathbb{E}(Y_2^{a_2=0} \mid A_2=0) + \mathbb{E}(Y_1 \mid A_2=0) \\
&\overset{\text{(A\ref{ass:consistency})}}{=} \mathbb{E}(Y_2 \mid A_2=1) - \mathbb{E}(Y_1 \mid A_2=1) - \mathbb{E}(Y_2 \mid A_2=0) + \mathbb{E}(Y_1 \mid A_2=0) \\
&= \left\{ \mathbb{E}(Y_2 \mid A_2=1) - \mathbb{E}(Y_1 \mid A_2=1) \right\} - \left\{ \mathbb{E}(Y_2 \mid A_2=0) - \mathbb{E}(Y_1 \mid A_2=0) \right\}.
\end{align*}
Assumption \ref{ass:positivity} guarantees that all conditional expectations are well-defined.

\section{}
\label{appendix:different_parallel_trend}
\begin{align*}
&\left\{ \mathbb{E}(Y_2 \mid A_2=1) - \mathbb{E}(Y_1 \mid A_2=1) \right\} - \left\{ \mathbb{E}(Y_2 \mid A_2=0) - \mathbb{E}(Y_1 \mid A_2=0) \right\}\\
&\overset{\text{(A\ref{ass:consistencyP})}}{=} \left\{ \mathbb{E}(Y_2 \mid A^{p=P}_2=1) - \mathbb{E}(Y_1 \mid A^{p=P}_2=1) \right\} - \left\{ \mathbb{E}(Y_2 \mid A^{p=P}_2=0) - \mathbb{E}(Y_1 \mid A^{p=P}_2=0) \right\} \\
&\overset{\text{(A\ref{ass:determinism2times})}}{=} \left\{ \mathbb{E}(Y_2 \mid P=1) - \mathbb{E}(Y_1 \mid P=1) \right\} - \left\{ \mathbb{E}(Y_2 \mid P=0) - \mathbb{E}(Y_1 \mid P=0) \right\} \\
&\overset{\text{(A\ref{ass:consistencyP})}}{=} \left\{ \mathbb{E}(Y^{p=1}_2 \mid P=1) - \mathbb{E}(Y^{p=1}_1 \mid P=1) \right\} - \left\{ \mathbb{E}(Y^{p=0}_2 \mid P=0) - \mathbb{E}(Y^{p=0}_1 \mid P=0) \right\} \\
&\overset{\text{(A\ref{ass:paralleltrendP})}}{=} \mathbb{E}(Y^{p=1}_2 \mid P=1) - \mathbb{E}(Y^{p=1}_1 \mid P=1) - \mathbb{E}(Y^{p=0}_2 \mid P=0) + \mathbb{E}(Y^{p=0}_1 \mid P=0) \\
&\quad - \mathbb{E}(Y_2^{p = 0} - Y_1^{p = 0} \mid P = 1) + \mathbb{E}(Y_2^{p = 0} - Y_1^{p = 0} \mid P = 0) \\
&= \mathbb{E}(Y^{p=1}_2 \mid P=1) - \mathbb{E}(Y^{p=1}_1 \mid P=1) - \mathbb{E}(Y_2^{p = 0} - Y_1^{p = 0} \mid P = 1) \\
&= \mathbb{E}(Y^{p = 1,\,a_2=A^{p=1}_2}_2 \mid P=1) - \mathbb{E}(Y^{p=1}_1 \mid P=1) - \mathbb{E}(Y_2^{p = 0,\,a_2=A^{p=0}_2} \mid P=1) + \mathbb{E}(Y_1^{p = 0} \mid P = 1) \\
&\overset{\text{(A\ref{ass:determinism2times})}}{=} \mathbb{E}(Y^{p = 1,\,a_2=1}_2 \mid P=1) - \mathbb{E}(Y^{p=1}_1 \mid P=1) - \mathbb{E}(Y_2^{p = 0,\,a_2=0} \mid P=1) + \mathbb{E}(Y_1^{p = 0} \mid P = 1) \\
&\overset{\text{(A\ref{ass:no_directeffectP})}}{=} \mathbb{E}(Y^{a_2=1}_2 \mid P=1) - \mathbb{E}(Y^{p=1}_1 \mid P=1) - \mathbb{E}(Y_2^{a_2=0} \mid P=1) + \mathbb{E}(Y_1^{p = 0} \mid P = 1) \\
&\overset{\text{(A\ref{ass:consistencyP})}}{=} \mathbb{E}(Y^{a_2=1}_2 \mid P=1,A_2=A^{p=P}_2) - \mathbb{E}(Y^{p=1}_1 \mid P=1) \\
&\quad - \mathbb{E}(Y_2^{a_2=0} \mid P=1,A_2=A^{p=P}_2) + \mathbb{E}(Y_1^{p = 0} \mid P = 1) \\
&\overset{\text{(A\ref{ass:determinism2times})}}{=} \mathbb{E}(Y^{a_2=1}_2 \mid A_2=1) - \mathbb{E}(Y^{p=1}_1 \mid P=1) - \mathbb{E}(Y_2^{a_2=0} \mid A_2=1) + \mathbb{E}(Y_1^{p = 0} \mid P = 1) \\
&\overset{\text{(D\ref{def:att_a2})}}{=} ATT_{A_2} - \{ \mathbb{E}(Y^{p=1}_1 \mid P=1) - \mathbb{E}(Y_1^{p = 0} \mid P = 1)\}.
\end{align*}
This proves Proposition \ref{prop:same_parallel_trend}, as $\psi= \mathbb{E}(Y^{p=1}_1 \mid P=1) - \mathbb{E}(Y_1^{p = 0} \mid P = 1)$.
Assumption \ref{ass:positivity} guarantees that all conditional expectations are well-defined.

\section{}
\label{appendix:cars}
Suppose that a car company unintentionally shipped cars with a defect to certain regions ($U = 1$). This defect increases the chances of having a car accident. Regions that received cars without the defect are denoted by $U = 0$. The moment the policymakers become aware of the problem, they decide to track down and seize all the interested cars. However, the seizing cannot take place immediately; the authorities need two days before they can get the full list of compromised vehicles and contact all the owners. Therefore, the defective cars are initially free to circulate ($A_1 = 0$), but two days after the decision all the defective vehicles are identified and seized ($A_2 = 1$). However, the fact that the cars are defective becomes public knowledge when the decision is made by the policymakers ($P = 1$). Thus, drivers of the defective cars, scared of being involved in accidents, decided to use replacement cars. This behavior fits with the plain English notion of ``anticipation'' in the DiD literature \citep{Malani2015} and with a violation of our Assumption \ref{ass:noanticipationP}.

Imagine that a researcher is interested in the effect of the policy on the number of car accidents, in the regions where the policy was introduced. To make a concrete numerical example corresponding to the story, consider the following structural causal model:
\begin{align*}
U &:\sim \text{Bernoulli}(0.2) \\
A_1 &:= 0 \\
P &:= U \\
Y_1 &:\sim \text{Binomial}\left(20, \frac{10 + 5 \cdot U - 5 \cdot P}{20}\right) \\
A_2 &:= P \\
Y_2 &:\sim \text{Binomial}\left(20, \frac{10 - 0.2 \cdot Y_1 + 5 \cdot U - 5 \cdot \max(P, A_2))}{20}\right)
\end{align*}
Here, $Y_1$ represents the number of car accidents in the region in the day after the policy decision and $Y_2$ represents the number of car accidents the following day, when the policy has actually been implemented. The number of accidents is affected by the presence of defective cars ($U = 1$). Moreover, if on the first day several accidents are reported on the news, the following day drivers will be more careful; hence the negative coefficient of $Y_1$ in the structural equation for $Y_2$.

After the decision of removing the defective cars ($P = 1$), all affected drivers ``anticipate'' the policy and completely offset the danger of the defective cars. This is why the coefficients for $U$ and $P$ are equal in magnitude but opposite in sign in the structural equation for $Y_1$. Similarly, the coefficients for $U$ and $A_2$ or $P$ offset each other in the structural equation for $Y_2$, as either the policy implementation or the spontaneous ``anticipation'' completely offset the danger. 

It is easy to see that Assumption \ref{ass:positivity} holds under this data generating mechanism. Also, the parallel trends in Assumption \ref{ass:parallel} holds because
\begin{align*}
\mathbb{E}(Y_2^{a_2 = 0} - Y_1 \mid A_2 = 1) 
&= \mathbb{E}(Y_2^{a_2 = 0} - Y_1 \mid P = 1, U = 1) \\
&= \mathbb{E}(Y_2^{a_2 = 0} \mid P = 1, U = 1) - \mathbb{E}(Y_1 \mid P = 1, U = 1) = 8 - 10 = -2
\end{align*}
\begin{align*}
\mathbb{E}(Y_2^{a_2 = 0} - Y_1 \mid A_2 = 0) 
&= \mathbb{E}(Y_2^{a_2 = 0} - Y_1 \mid P = 0, U = 0) \\
&= \mathbb{E}(Y_2^{a_2 = 0} \mid P = 0, U = 0) - \mathbb{E}(Y_1 \mid P = 0, U = 0) = 8 - 10 = -2.
\end{align*}

Therefore, it is possible to identify $ATT_{A_2}$ using the functional in Equation \ref{eq:classicdid}, which in this case is equal to 0:
\begin{align*}
ATT_{A_2} &\overset{\text{(D\ref{def:att_a2})}}{=} \mathbb{E}(Y_2^{a_2 = 1} - Y_2^{a_2 = 0} \mid A_2 = 1) \\
&= \mathbb{E}(Y_2^{a_2 = 1} \mid P = 1, U = 1) - \mathbb{E}(Y_2^{a_2 = 0} \mid P = 1, U = 1) = 8 - 8 = 0
\end{align*}
and indeed the classic DiD functional is
\begin{align*}
\left\{ \mathbb{E}(Y_2 \mid A_2 = 1) - \mathbb{E}(Y_1 \mid A_2 = 1) \right\} - \left\{ \mathbb{E}(Y_2 \mid A_2 = 0) - \mathbb{E}(Y_1 \mid A_2 = 0) \right\}=-2-(-2)=0. 
\end{align*}
Under this data generating mechanism, the expected number of accidents for regions with $A_2 = 1$ does not change if we intervene on $A_2$. The reason is that $P=1$ whenever we observe $A_2=1$ and in regions with $P = 1$ there is no effect of $A_2$ on $Y_2$.

The \textit{implementation} of the policy, that is, intervening on $A_2$, has no effect in these regions. However, the \textit{decision of implementing} the policy has a beneficial effect: the $ATT_P$ is equal to $-4$,
\begin{align*}
ATT_P &\overset{\text{(D\ref{def:att_p})}}{=} \mathbb{E}(Y_2^{p = 1} - Y_2^{p = 0} \mid P = 1) = \mathbb{E}(Y_2^{p = 1} \mid U = 1) - \mathbb{E}(Y_2^{p = 0} \mid U = 1) = 8 - 12 = -4.
\end{align*}
The decision of stopping the defective vehicles lowered the number of car accidents.

\section{}
\label{appendix:didwithp_2times}

In Appendix \ref{appendix:different_parallel_trend}, we have shown that under assumptions \ref{ass:positivity}, \ref{ass:determinism2times}, \ref{ass:paralleltrendP}, \ref{ass:consistencyP}
\begin{align*}
&\left\{ \mathbb{E}(Y_2 \mid A_2=1) - \mathbb{E}(Y_1 \mid A_2=1) \right\} - \left\{ \mathbb{E}(Y_2 \mid A_2=0) - \mathbb{E}(Y_1 \mid A_2=0) \right\}\\
&= \mathbb{E}(Y^{p=1}_2 \mid P=1) - \mathbb{E}(Y^{p=1}_1 \mid P=1) - \mathbb{E}(Y_2^{p = 0} - Y_1^{p = 0} \mid P = 1).
\end{align*}
Proposition \ref{prop:didwithp_2times} is immediately proven by recognizing that
\begin{align*}
&\mathbb{E}(Y^{p=1}_2 \mid P=1) - \mathbb{E}(Y^{p=1}_1 \mid P=1) - \mathbb{E}(Y_2^{p = 0} - Y_1^{p = 0} \mid P = 1)\\
&=\mathbb{E}(Y^{p=1}_2 \mid P=1) - \mathbb{E}(Y^{p=1}_1 \mid P=1) - \mathbb{E}(Y_2^{p = 0} \mid P = 1) + \mathbb{E}(Y_1^{p = 0} \mid P = 1)\\
&\overset{\text{(D\ref{def:att_p})}}{=} ATT_P - \mathbb{E}(Y^{p=1}_1 \mid P=1) + \mathbb{E}(Y_1^{p = 0} \mid P = 1) \overset{\text{(A\ref{ass:noanticipationP})}}{=} ATT_P.
\end{align*}

\section{}
\label{appendix:didwithp_multtime}

Consider a binary variable $P_k$ indicating whether the policy makers have decided, at time $k$ or before, to implement ($P_k = 1$) or not ($P_k = 0$) the policy. Once the decision is first made, this variable will always take the value 1. We assume a FFRCISTG model where at each time $k$, the decision $P_k$ precedes the outcome $Y_k$.

At the beginning of the study, all units are untreated and no units planned to implement the policy before time 1. However, when the decision of implementing the policy is made, the policy enters into force after $s \geq 1$ time points and remains in place. For simplicity, we assume that the treatment implementation indicators $A_k$ are defined until $\tau+s$.

\begin{assumption}
\label{ass:determinism_multtime}
Suppose that $A_k = 0$ for all $k$ such that $k \leq s$. For all $k \in \{s+1, \ldots, \tau+s\}$,
\[
A_k = P_{k-s}.
\]
\end{assumption}

Assumption~\ref{ass:determinism_multtime} only enforces a determinism between decision and lagged implementation in the data. Since we are interested in the effect of interventions on $P$, this assumption will suffice for our identification argument.

We indicate with $\bar{P}_k$ the history of all measurements of this variable from time 1 to time $k$. We will assume that there is a positive probability of first making the decision to implement the policy up to time $\tau - s$. For simplicity, we will assume that if, by time $\tau - s$, a unit decided to not implement the policy, it will persist in the decision to not implement the policy until the end of the study. This restriction is motivated by the fact that in some settings the decision variable is not directly observed, and it is not possible to use $\bar{A}_\tau$ to infer the value of $P_k$ for units with $P_{\tau-s} = 0$ and times $\tau - s < k \leq \tau$.

\begin{assumption}
\label{ass:decision_structure}
In the set 
\[
\mathcal{P} = \left\{ \bar{p}_\tau \in \{0,1\}^\tau \mid \exists h \in \{0, \ldots, \tau - s\} \text{ such that } p_k = 
\begin{cases}
0 & \text{when } 1 \leq k \leq h \\
1 & \text{when } h < k \leq \tau - s \\
p_{\tau - s} & \text{when } \tau - s < k \leq \tau
\end{cases}
\right\},
\]
we have that $\Pr(\bar{P}_\tau = \bar{p}_\tau) > 0 \iff \bar{p}_\tau \in \mathcal{P}$.
\end{assumption}

Similarly to the approach described in Section ``\nameref{sec:different_intervention}'', we conceptualize an intervention on the variables $P_k$ and use the potential outcome notation $Y_k^{\bar{p}_\tau}$ to indicate interventions on the vector of values of $P_k$ until time $\tau$. This motivates a new estimand for $g \leq \tau - s$ and $k \leq \tau$:

\begin{definition}
\label{def:att_Pmulttime}
\[
ATT_{\bar{P}}(g,k) = \mathbb{E}\left( Y_k^{\bar{p}_\tau = \{ \bar{0}_{g-1}, \bar{1}_{\tau - g + 1} \}} - Y_k^{\bar{p}_\tau = \bar{0}_\tau} \mid \bar{P}_\tau = \{ \bar{0}_{g-1}, \bar{1}_{\tau - g + 1} \} \right).
\]
\end{definition}

The $ATT_{\bar{P}}(g,k)$ represents the effect, at time $k$, of first deciding at time $g$ to implement the policy versus deciding during the study period to not implement the policy, among units that first decided to implement the policy at time $g$. This estimand is conceptually similar to the ``group-time average treatment effect'' proposed by \citet{callaway2021did,callaway2024multiperoid} reported in Definition \ref{def:att_Amulttime}, except that we are now considering a different intervention, which is the intervention on the decision rather than on the policy implementation.

We will also invoke a consistency assumption, ensuring correspondence between potential outcomes and observed outcomes.
\begin{assumption}
\label{ass:consistencyP_multtime}
For every $k \in \{1, \ldots, \tau\}$ and $\bar{p}_\tau \in \mathcal{P}$, $Y_k^{\bar{p}_\tau} = Y_k$ when $\bar{P}_\tau = \bar{p}_\tau$.
\end{assumption}

Furthermore, we will invoke the following no-anticipation assumption.
\begin{assumption}
\label{ass:noanticipation_multtime}
For every $g \leq \tau - s$, and $g \leq k < g + s$:
\[
\mathbb{E}\left( Y_k^{\bar{p}_\tau = \{ \bar{0}_{g-1}, \bar{1}_{\tau - g + 1} \}} \mid \bar{P}_\tau = \{ \bar{0}_{g-1}, \bar{1}_{\tau - g + 1} \} \right) = \mathbb{E}\left( Y_k^{\bar{p}_\tau = \bar{0}_\tau} \mid \bar{P}_\tau = \{ \bar{0}_{g-1}, \bar{1}_{\tau - g + 1} \} \right).
\]
\end{assumption}

Assumption \ref{ass:noanticipation_multtime} requires that, among units deciding at time $g$ to implement the policy, there is no difference in the expected outcomes between the treatment strategy “decide at time g to first implement the policy” and “decide during the study to not implement the policy”, as long as $k<g+s$. Informally, this means that the decision to implement the policy at time $g$ does not affect the outcomes until the implementation actually takes place ($k=g+s$). A trivial implication of Assumption \ref{ass:noanticipation_multtime} is that for $g \leq \tau - s$, and $k \leq \tau$ such that $k<g+s$, $ATT_{\bar{P}}(g,k)=0$.

We will finally invoke a parallel trends assumption between units that decided at time $g$ to implement the policy and units that decided to not implement the policy until later.
\begin{assumption}
\label{ass:paralleltrend_multtime}
For every $g \leq \tau - s$, every $k \geq g + s$ and every $k-s \leq j \leq \tau$
\[
\mathbb{E}(Y_k^{\bar{p}_\tau = \bar{0}_\tau} \mid \bar{P}_\tau = \{ \bar{0}_{g-1}, \bar{1}_{\tau - g + 1} \}) - \mathbb{E}(Y_{k-1}^{\bar{p}_\tau = \bar{0}_\tau} \mid \bar{P}_\tau = \{ \bar{0}_{g-1}, \bar{1}_{\tau - g + 1} \}) 
\]
\[
= \mathbb{E}(Y_k^{\bar{p}_\tau = \bar{0}_\tau} \mid \bar{P}_j = \bar{0}_j) - \mathbb{E}(Y_{k-1}^{\bar{p}_\tau = \bar{0}_\tau} \mid \bar{P}_j = \bar{0}_j).
\]
\end{assumption}

Here, we consider the parallel trends to hold with all units who decided not to implement the policy by time $j$, for every $j$ between $k-s$ and the end of the study. By narrowing the range for $j$, it is possible to reduce Assumption \ref{ass:paralleltrend_multtime} to either of the parallel trends assumptions in \citet{callaway2021did} using the decision as the treatment of interest.
Under these assumptions, the $ATT_{\bar{P}}(g,k)$ is identified by a functional as the one proposed by \citet{callaway2021did}.
\begin{proposition}
\label{prop:didwithp_multtime}
Suppose that Assumptions~\ref{ass:determinism_multtime}–\ref{ass:paralleltrend_multtime} hold. For every $g \leq \tau - s$ and every $k \geq g + s$, 
\begin{align*}
ATT_{\bar{P}}(g,k) &= \left\{ \mathbb{E}(Y_k \mid A_{g+s} = 1, \bar{A}_{g+s - 1} = \bar{0}_{g+s - 1}) - \mathbb{E}(Y_{g+s - 1} \mid A_{g+s} = 1, \bar{A}_{g+s - 1} = \bar{0}_{g+s - 1}) \right\} \\
&\quad - \left\{ \mathbb{E}(Y_k \mid \bar{A}_{k} = \bar{0}_{k}) - \mathbb{E}(Y_{g+s - 1} \mid \bar{A}_{k} = \bar{0}_{k}) \right\}\\
&= \left\{ \mathbb{E}(Y_k \mid A_{g+s} = 1, \bar{A}_{g+s - 1} = \bar{0}_{g+s - 1}) - \mathbb{E}(Y_{g+s - 1} \mid A_{g+s} = 1, \bar{A}_{g+s - 1} = \bar{0}_{g+s - 1}) \right\} \\
&\quad - \left\{ \mathbb{E}(Y_k \mid \bar{A}_\tau = \bar{0}_\tau) - \mathbb{E}(Y_{g+s - 1} \mid \bar{A}_\tau = \bar{0}_\tau) \right\}.
\end{align*}
\end{proposition}
See below for the proof.
This result is a direct generalization of the scenario discussed in Section ``\nameref{sec:different_intervention}'', where $\tau=2$ and $s=1$. In this setting, the $ATT_{\bar{P}}(1,2)$ corresponds to the functional in Proposition \ref{prop:didwithp_2times}.

The functionals in Proposition~\ref{prop:didwithp_multtime} are identical to the functionals from \citet{callaway2021did,callaway2024multiperoid} when considering an intervention at time $g+s$ rather than $g$; the $ATT_{\text{unc}}^{\text{ny}}(g+s,k;\delta=0)$ and the $ATT_{\text{unc}}^{\text{nev}}(g+s,k;\delta=0)$. \citet{callaway2021did,callaway2024multiperoid} also allow $\delta > 0$ to indicate a degree of ``anticipation'', see the Standard limited treatment anticipation assumption. In our setup this would correspond to allowing Assumption~\ref{ass:noanticipation_multtime} to hold only for $k$ such that $g \leq k < g + s - \delta$. As the authors point out, however, this then requires the parallel trends assumption to be true also for time points before $g + s$~\citep{callaway2021did}.

\vspace{0.5cm}
\noindent{\textit{Proof of Proposition \ref{prop:didwithp_multtime}}}
For every \( g \leq \tau - s \), every \( k \geq g + s \), and every \( k-s \leq j \leq \tau \),
\begin{align*}
&\text{ATT}_{\bar{P}}(g,k)\overset{\text{(D\ref{def:att_Pmulttime})}}{=} \mathbb{E}\left( Y_k^{\bar{p}_\tau = \{ \bar{0}_{g-1}, \bar{1}_{\tau - g + 1} \}} -  Y_k^{\bar{p}_\tau = \bar{0}_\tau} \mid \bar{P}_\tau = \{ \bar{0}_{g-1}, \bar{1}_{\tau - g + 1} \} \right) \\
&\overset{\text{(A\ref{ass:paralleltrend_multtime})}}{=} \mathbb{E}\left( Y_k^{\bar{p}_\tau = \{ \bar{0}_{g-1}, \bar{1}_{\tau - g + 1} \}} \mid \bar{P}_\tau = \{ \bar{0}_{g-1}, \bar{1}_{\tau - g + 1} \} \right) - \mathbb{E}\left( Y_k^{\bar{p}_\tau = \bar{0}_\tau} \mid \bar{P}_\tau = \{ \bar{0}_{g-1}, \bar{1}_{\tau - g + 1} \} \right) \\
&\quad - \mathbb{E}\left( Y_k^{\bar{p}_\tau = \bar{0}_\tau} \mid \bar{P}_j = \bar{0}_j \right) + \mathbb{E}\left( Y_{k-1}^{\bar{p}_\tau = \bar{0}_\tau} \mid \bar{P}_j = \bar{0}_j \right) \\
&\quad + \mathbb{E}\left( Y_k^{\bar{p}_\tau = \bar{0}_\tau} \mid \bar{P}_\tau = \{ \bar{0}_{g-1}, \bar{1}_{\tau - g + 1} \} \right) - \mathbb{E}\left( Y_{k-1}^{\bar{p}_\tau = \bar{0}_\tau} \mid \bar{P}_\tau = \{ \bar{0}_{g-1}, \bar{1}_{\tau - g + 1} \} \right) \\
&= \mathbb{E}\left( Y_k^{\bar{p}_\tau = \{ \bar{0}_{g-1}, \bar{1}_{\tau - g + 1} \}} \mid \bar{P}_\tau = \{ \bar{0}_{g-1}, \bar{1}_{\tau - g + 1} \} \right) - \mathbb{E}\left( Y_k^{\bar{p}_\tau = \bar{0}_\tau} \mid \bar{P}_j = \bar{0}_j \right) \\
&\quad + \mathbb{E}\left( Y_{k-1}^{\bar{p}_\tau = \bar{0}_\tau} \mid \bar{P}_j = \bar{0}_j \right) - \mathbb{E}\left( Y_{k-1}^{\bar{p}_\tau = \bar{0}_\tau} \mid \bar{P}_\tau = \{ \bar{0}_{g-1}, \bar{1}_{\tau - g + 1} \} \right)
\end{align*}
when $k>g+s$, we proceed iteratively
\begin{align*}
&\overset{\text{(A\ref{ass:paralleltrend_multtime})}}{=}
\mathbb{E}\left(Y_k^{\bar{p}_\tau = \{ \bar{0}_{g-1}, \bar{1}_{\tau - g + 1} \}} \mid \bar{P}_\tau = \{ \bar{0}_{g-1}, \bar{1}_{\tau - g + 1} \} \right)
- \mathbb{E}\left(Y_k^{\bar{p}_\tau = \bar{0}_\tau} \mid \bar{P}_j = \bar{0}_j \right) + \mathbb{E}\left(Y_{k-1}^{\bar{p}_\tau = \bar{0}_\tau} \mid \bar{P}_j = \bar{0}_j \right)\\
&\quad - \mathbb{E}\left(Y_{k-1}^{\bar{p}_\tau = \bar{0}_\tau} \mid \bar{P}_\tau = \{ \bar{0}_{g-1}, \bar{1}_{\tau - g + 1} \} \right) - \mathbb{E}\left(Y_{k-1}^{\bar{p}_\tau = \bar{0}_\tau} \mid \bar{P}_j = \bar{0}_j \right) + \mathbb{E}\left(Y_{k-2}^{\bar{p}_\tau = \bar{0}_\tau} \mid \bar{P}_j = \bar{0}_j \right)\\
&\quad + \mathbb{E}\left(Y_{k-1}^{\bar{p}_\tau = \bar{0}_\tau} \mid \bar{P}_\tau = \{ \bar{0}_{g-1}, \bar{1}_{\tau - g + 1} \} \right)
- \mathbb{E}\left(Y_{k-2}^{\bar{p}_\tau = \bar{0}_\tau} \mid \bar{P}_\tau = \{ \bar{0}_{g-1}, \bar{1}_{\tau - g + 1} \} \right) \\
&= 
\mathbb{E}\left(Y_k^{\bar{p}_\tau = \{ \bar{0}_{g-1}, \bar{1}_{\tau - g + 1} \}} \mid \bar{P}_\tau = \{ \bar{0}_{g-1}, \bar{1}_{\tau - g + 1} \} \right)
- \mathbb{E}\left(Y_k^{\bar{p}_\tau = \bar{0}_\tau} \mid \bar{P}_j = \bar{0}_j \right)\\
&\quad + \mathbb{E}\left(Y_{k-2}^{\bar{p}_\tau = \bar{0}_\tau} \mid \bar{P}_j = \bar{0}_j \right)
- \mathbb{E}\left(Y_{k-2}^{\bar{p}_\tau = \bar{0}_\tau} \mid \bar{P}_\tau = \{ \bar{0}_{g-1}, \bar{1}_{\tau - g + 1} \} \right)
\end{align*}
until we obtain
\begin{align*}
&=
\mathbb{E}\left( Y_k^{\bar{p}_\tau = \{ \bar{0}_{g-1}, \bar{1}_{\tau - g + 1} \}} \mid \bar{P}_\tau = \{ \bar{0}_{g-1}, \bar{1}_{\tau - g + 1} \} \right)
- \mathbb{E}\left( Y_k^{\bar{p}_\tau = \bar{0}_\tau} \mid \bar{P}_j = \bar{0}_j \right) \\
&\quad + \mathbb{E}\left( Y_{g+s-1}^{\bar{p}_\tau = \bar{0}_\tau} \mid \bar{P}_j = \bar{0}_j \right)
- \mathbb{E}\left( Y_{g+s-1}^{\bar{p}_\tau = \bar{0}_\tau} \mid \bar{P}_\tau = \{ \bar{0}_{g-1}, \bar{1}_{\tau - g + 1} \} \right) \\
&\overset{\text{(A\ref{ass:noanticipation_multtime})}}{=}
\mathbb{E}\left( Y_k^{\bar{p}_\tau = \{ \bar{0}_{g-1}, \bar{1}_{\tau - g + 1} \}} \mid \bar{P}_\tau = \{ \bar{0}_{g-1}, \bar{1}_{\tau - g + 1} \} \right)
- \mathbb{E}\left( Y_k^{\bar{p}_\tau = \bar{0}_\tau} \mid \bar{P}_j = \bar{0}_j \right) \\
&\quad + \mathbb{E}\left( Y_{g+s-1}^{\bar{p}_\tau = \bar{0}_\tau} \mid \bar{P}_j = \bar{0}_j \right)
- \mathbb{E}\left( Y_{g+s-1}^{\bar{p}_\tau = \{ \bar{0}_{g-1}, \bar{1}_{\tau - g + 1} \}} \mid \bar{P}_\tau = \{ \bar{0}_{g-1}, \bar{1}_{\tau - g + 1} \} \right) \\
&\overset{\text{(A\ref{ass:consistencyP_multtime},\ref{ass:noanticipation_multtime})}}{=}
\mathbb{E}\left( Y_k \mid \bar{P}_\tau = \{ \bar{0}_{g-1}, \bar{1}_{\tau - g + 1} \} \right)
- \mathbb{E}\left( Y_k \mid \bar{P}_j = \bar{0}_j \right)\\
&\quad + \mathbb{E}\left( Y_{g+s-1} \mid \bar{P}_j = \bar{0}_j \right)
- \mathbb{E}\left( Y_{g+s-1} \mid \bar{P}_\tau = \{ \bar{0}_{g-1}, \bar{1}_{\tau - g + 1} \} \right)\\
&\overset{\text{(A\ref{ass:determinism_multtime})}}{=}
\mathbb{E}\left( Y_k \mid A_{g+s} = 1, \bar{A}_{g+s-1} = \bar{0}_{g+s-1} \right)
- \mathbb{E}\left( Y_k \mid \bar{A}_{j+s} = \bar{0}_{j+s} \right) \\
&\quad + \mathbb{E}\left( Y_{g+s-1} \mid \bar{A}_{j+s} = \bar{0}_{j+s} \right)
- \mathbb{E}\left( Y_{g+s-1} \mid A_{g+s} = 1, \bar{A}_{g+s-1} = \bar{0}_{g+s-1} \right) \\
&=
\left\{ \mathbb{E}\left( Y_k \mid A_{g+s} = 1, \bar{A}_{g+s-1} = \bar{0}_{g+s-1} \right)
- \mathbb{E}\left( Y_{g+s-1} \mid A_{g+s} = 1, \bar{A}_{g+s-1} = \bar{0}_{g+s-1} \right) \right\} \\
&\quad -
\left\{ \mathbb{E}\left( Y_k \mid \bar{A}_{j+s} = \bar{0}_{j+s} \right)
- \mathbb{E}\left( Y_{g+s-1} \mid \bar{A}_{j+s} = \bar{0}_{j+s} \right) \right\}.
\end{align*}
Assumption \ref{ass:decision_structure} guarantees that all conditional expectations are well-defined. Choosing $j = k-s$ proves the first equality in Proposition \ref{prop:didwithp_multtime}. Choosing $j = \tau$ proves the second equality in Proposition \ref{prop:didwithp_multtime} because $\bar{A}_{\tau+s} = \bar{0}_{\tau+s} \overset{\text{(A\ref{ass:determinism_multtime})}}{\iff} P_\tau = 0 \overset{\text{(A\ref{ass:decision_structure})}}{\iff} P_{\tau-s} = 0 \overset{\text{(A\ref{ass:determinism_multtime})}}{\iff} \bar{A}_\tau = \bar{0}_\tau$. 
\end{document}